\documentclass[preprint,showpacs,preprintnumbers,amsmath,amssymb,natbib]{revtex4}

\usepackage{graphicx}
\usepackage{epsfig}		
\usepackage{dcolumn}
\usepackage{bm}

\def\0{\mbox{\tiny $0$}}
\def\1{\mbox{\tiny $1$}}
\def\2{\mbox{\tiny $2$}}
\def\3{\mbox{\tiny $3$}}
\def\4{\mbox{\tiny $4$}}
\def\5{\mbox{\tiny $5$}}
\def\6{\mbox{\tiny $6$}}
\def\7{\mbox{\tiny $7$}}
\def\8{\mbox{\tiny $8$}}
\def\9{\mbox{\tiny $9$}}

\def\f14{\mbox{\tiny $\frac{1}{4}$}}

\begin{document}

\title{$SU(2) \otimes SU(2)$ bi-spinor structure entanglement induced by a step potential barrier scattering in two-dimensions}

\author{Victor A. S. V. Bittencourt}
\email{vbittencourt@df.ufscar.br}
\author{Salomon S. Mizrahi}
\email{salomon@df.ufscar.br}
\author{Alex E. Bernardini}
\email{alexeb@ufscar.br}
\affiliation{Departamento de F\'{\i}sica, Universidade Federal de S\~ao Carlos, PO Box 676, 13565-905, S\~ao Carlos, SP, Brasil}

\date{\today}

\begin{abstract}
The entanglement between $SU(2) \otimes SU(2)$ internal degrees of freedom of parity and helicity for reflected and transmitted waves of Dirac-like particles scattered by a potential step along an arbitrary direction on the $x-y$ plane is quantified. Diffusion ($E \geq V$) and Klein zone  ($V \geq E$) energy regimes are considered. It has been shown that, for $SU(2) \otimes SU(2)$ polarized structures of helicity eigenstates impinging the barrier, the local interaction with a {\em step} potential destroys the \textit{parity-spin} separability.
The framework presented here can be straightforwardly translated into a useful theoretical tool for obtaining the \textit{spin-spin} entanglement in the context of enlarged scenarios of nonrelativistic $2D$ systems, as for instance those for describing single layer graphene, or even single trapped ions with Dirac bi-spinor mathematical structure.
\end{abstract}


\date{\today}
\maketitle

\section{Introduction}

The growing interest in mapping controllable physical systems onto the bi-spinor structure of the relativistic Dirac equation \cite{n001,n002,n003,n004} has been recently translated into a useful theoretical tool for investigating nonrelativistic two-dimensional ($2D$) systems of utmost scientific appeal \cite{n005,n006,n007}.
For instance, the low-energy excitations of a nonrelativistic electrons in the single layer graphene are known for exhibiting a massless Weyl spinor structure often related to $2D$ Dirac equation solutions.
Likewise, on the frontier of the relativistic quantum mechanics connections to the solid state physics, the blackhole-like properties in Bose-Einstein condensates \cite{n008}, the simulation of Unruh effect in trapped ions \cite{n009}, the {\em trembling} motion and the Klein's paradox for massive fermions in $2D$ systems \cite{n010} have all been currently investigated in both theoretical and experimental scopes.

On the other hand, quantum information and special relativity have composed the intermediate framework for identifying the entanglement/separability information content of Dirac bi-spinors, even if the Lorentz symmetry is assumed \cite{n011}.
A recent assertive result \cite{n012,003} establishes that the $SU(2) \otimes SU(2)$ group structure of Dirac bi-spinors are assigned to a Dirac Hamiltonian written in terms of the direct product of two-qubit operators, ${H}_{D}={\sigma}_{x}^{\left( 1\right) }\otimes \left(
\vec{p}\cdot \vec{{\sigma}}^{\left( 2\right) }\right) +m \,
{\sigma }_{z}^{\left( 1\right)}\otimes {I}^{(2)}_{2}$,
from which the free particle solutions of the Dirac equation are given in terms of $SU(2) \otimes SU(2)$ \textit{parity-spin} entangled states written as
\begin{eqnarray}
\left\vert \Psi ^{s}(\vec{p},\,t)\right\rangle
&=&e^{i(-1)^{s}\,E_{p}\,t}\left\vert \psi ^{s}(\vec{p})\right\rangle
= e^{i(-1)^{s}\,E_{p}\,t}N_{s}\left( p\right)  \notag
\\
&\times& \left[ \left\vert
+\right\rangle _{1}\otimes \left\vert u(\vec{p})\right\rangle _{2}+\left(
\frac{p}{E_{p}+(-1)^{s+1}m}\right) |-\rangle _{1}\,\otimes \left( \hat{p}
\cdot \vec{{\sigma }}^{\left( 2\right) }\left\vert u(\vec{p}
)\right\rangle _{2}\right) \right], \label{sol1}
\end{eqnarray}
where $s=0$ and $1$ stand respectively for negative and positive frequencies, and $\vec{p} = p \hat{p}$ with $\hat{p}$ as a unitary vector.
The spinor described by $u(\vec{p})$ in the momentum representation is related to the spatial motion of the respective Dirac solution, which is coupled to its spin and describes a structureless magnetic dipole moment \cite{003}.
For the first qubit, the kets, $\vert + \rangle$ and $\vert-\rangle$, are categorized as the mutually orthogonal intrinsic parity eigenstates of the spinorial particle, and the corresponding total inner product set as $\vert \Psi ^{s}(\vec{p},\,t)\vert^{2} = \vert u(\vec{p})\vert^{\2}$ is supported by a normalization factor given by
$$N_{s}(p) = \frac{1}{\sqrt{2}} \left[1+ (1)^{s+1}\frac{m}{E_p}\right]^{\frac{1}{2}}.$$
Such a state vector Dirac solution, $\Psi ^{s}(\vec{p},\,t)$ is also used to be written as the sum of two bi-spinor structures in the momentum representation given by
\begin{eqnarray}
\Psi ^{s}(\vec{p},\,t)
&=& e^{i(-1)^{s}\,E_{p}\,t} \hspace{-.1 cm} N_{s}(p) \left[ \begin{array}{c} u(\vec{p}) \\ \frac{\vec{p} \cdot \vec{\sigma}}{E_p + (-1)^{s + 1} m} u(\vec{p}) \end{array} \right],
\end{eqnarray}
for which, a most complete description of its intrinsic quantum correlation properties are given in Ref.\cite{003}.
Its corresponding explicit \textit{parity-spin} entanglement structure can be straightforwardly extended to simulate all the above preliminarily mentioned physical systems since their similar Dirac bi-spinor structures may exhibit some analogous {\em spin-spin} entanglement, or even additional quantum correlations.

Quantifying the entanglement becomes more interesting whether one considers that the simplest and mostly known scheme for introducing external interaction elements is the {\em step} potential barrier problem. Even locally, it may affect the dynamical evolution of any Dirac bi-spinor structure.
As it shall be shown, for $SU(2) \otimes SU(2)$ structures prepared as separable helicity eigenstates, the local interaction with a {\em step} potential barrier destroys the \textit{parity-spin} separability such that the induced entanglement of scattered states can be exactly quantified.

Some preliminary issues have already investigated the relativistic linear transmission \cite{n013,Alex1,Alex4,Alex5} and the phenomena of planar diffusion of Dirac particles \cite{n014, n015,n016} through and above potential barriers and steps.
In particular, the effects of relative phases between incoming and reflected or transmitted amplitude components of diffused waves have been accurately quantified \cite{Alex2,Alex3,n013}.
Once that the Dirac equation solutions can be interpreted as two-qubit states -  one associated to the intrinsic parity and another to the helicity quantum number - the projection of particle's spin on its momentum direction - the corresponding \textit{parity-spin} entanglement associated to reflected and transmitted waves along the potential barrier can be systematically quantified.

Our aim is, therefore, to quantify the entanglement between intrinsic parity and helicity for reflected and transmitted waves of Dirac particles scattered by a potential step-barrier.
The impinging particles are assumed to move along a collinear arbitrary direction on the $x-y$ plane.
Given that incident waves can be prepared as separable quantum superpositions of positive and negative helicity eigenstates, the \textit{parity-spin} entanglement for reflected and transmitted bi-spinors shall be determined through the corresponding entanglement quantifier.
In case of pure states, it is computed through the von-Neumann (vN) quantum entropy which is driven by the $x-y$ incidence angle, $\theta$, and by the relative phase between the incident amplitudes.

The paper is organized as follows.
In Sec. II, the Dirac solutions for the step potential barrier problem in diffusion and Klein (paradox) regimes are obtained, with the corresponding amplitudes of probability for reflected and transmitted waves given in terms of incidence and transmission angles, $\theta\,$ and $\theta^{\prime\,}$.
In Sec. III, the \textit{parity-spin} entanglement for reflected and transmitted waves described as pure states is computed in terms of the vN quantum entropy.
The transition between scattering (diffusion and Klein zone \cite{n018,n019}) and barrier penetration regimes is identified by the entanglement profile, as well as through the profile of the averaged chirality of transmitted and reflected waves.
The relativistic limits are obtained, and the possibility of including phase differences between helicity degrees of freedom (DoF) is also investigated.
Our concluding remarks are given in Sec. IV, where lessons concerning the extension of our results to Dirac-like systems exhibiting {\em spin-spin} analogous to {\em parity-spin} entanglement are drawn.

\section{Dirac solutions for the step potential barrier}

The Dirac Hamiltonian for a free bi-spinor structure physically realized by, for instance, an electron, is given by
\begin{equation}
H_D  = - i \vec{\alpha} \cdot \vec{\nabla} + \beta m^2,
\label{Ham}
\end{equation}
with the $4 \times 4$ matrices $\vec{\alpha} \equiv \{\alpha_x, \alpha_y, \alpha_z \}$ and $\beta$ satisfying the algebra relations,
\begin{equation}
\alpha_k \alpha_l + \alpha_l \alpha_k = 2 I \delta_{kl}, \hspace{ 0.9 cm} \vec{\alpha} \beta + \beta \vec{\alpha} = 0, \hspace{ 0.9 cm} \beta^2 = I.
\end{equation}
In the Pauli-Dirac representation the matrices $\vec{\alpha}$ and $\beta$ are written in terms of tensor products of Pauli matrices acting on two subspaces labeled $1$ and $2$ as
\begin{equation}
\vec{\alpha} \cdot \vec{\nabla} = \sigma_x^{(1)} \otimes (\vec{\nabla} \cdot \vec{\sigma}^{(2)}), \hspace{0.9 cm} \beta = \sigma_z ^{(1)} \otimes I^{(2)}_2,
\end{equation}
such that one could interpret the solutions of Dirac equation as a state of two-qubits carried by a massive particle whose dynamical evolution is set up by the Hamiltonian Eq.~(\ref{Ham}) \cite{n012, n017}.
Noticing that all the calculations for the step barrier scattering problem (which includes the particle's diffusion and the Klein zone regimes) can be carried out in the momentum space, the positive energy plane wave solutions of Dirac equation, $\vert \psi_\pm \rangle$, obtained from (\ref{sol1}), can be simply described as eigenstates of the helicity operator $h = \vec{\Sigma} \cdot \vec{p} / p$ \cite{n018}, with eigenvalues $ \pm 1$, given by
\begin{equation}
\label{001}
\vert \psi_\pm (\vec{x},t) \rangle =\sqrt{\frac{E - m}{2 E}} \, \left[\,  \vert 1 \rangle \otimes \vert h_\pm \rangle \pm \sqrt{\frac{E + m}{E-m}} \vert 0 \rangle \otimes \vert h_\pm \rangle \, \right]\, \exp[- i (E t - \vec{p} \cdot \vec{x})],
\end{equation}
with
\begin{equation}
\label{e001}
\vert h_{\pm} \rangle = \frac{1 \pm \vec{\sigma}\cdot \hat{p}}{\sqrt{2}} \, \vert \pm \rangle,
\end{equation}
and where the kets $\vert 1 \rangle$ and $\vert 0 \rangle$ carry the intrinsic parity quantum number, as to have $P \vert p \rangle = (-1)^{p}\vert p \rangle$, with $p = 1$ and $0$ respectively for {\em odd} and {\em even} intrinsic parity, and the kets $\vert h_{\pm} \rangle$ are given in terms of $\sigma_z ^{(2)}$ eigenstates $\vert \pm \rangle$ through relation (\ref{e001}).

Considering that interactions with electromagnetic fields could be included in the Dirac equation, the presence of an electrostatic constant potential leads to the non-covariant form of the Dirac equation as \cite{n018}
\begin{equation}
i \frac{\partial \psi}{\partial t} = (H_D + V) \psi,
\end{equation}
from which stationary solutions of $H_D \psi = (E - V) \psi$, as well as their corresponding bi-spinor structure, could be straightforwardly obtained through the substitutions $E \rightarrow E - V$ and $p \rightarrow  \sqrt{(E- V)^2 - m^2} = q$ into Eq.~(\ref{001}).

\begin{figure}[b]
\includegraphics[width = 10 cm]{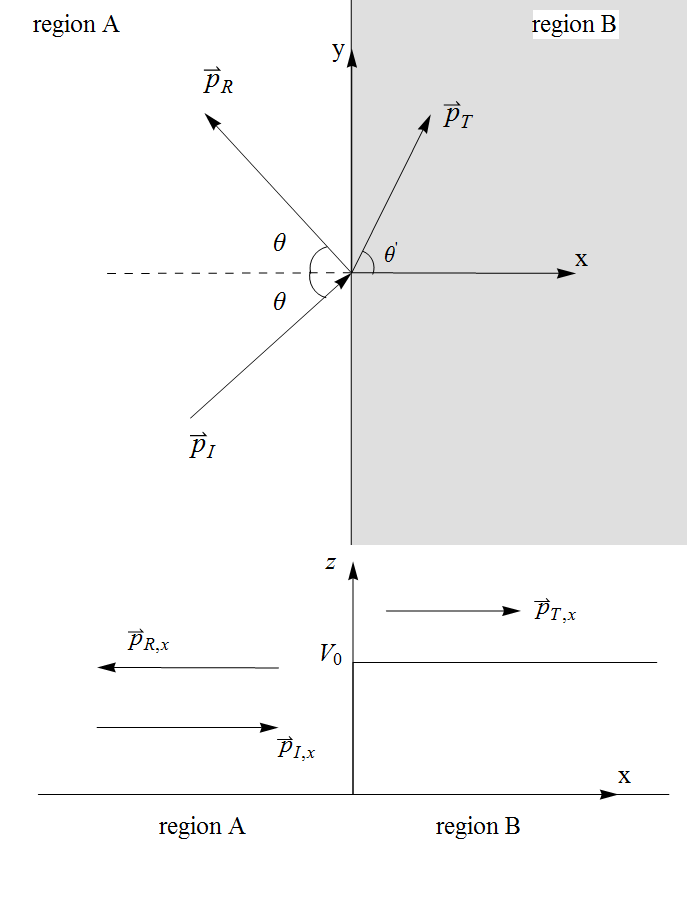}
\caption{Schematic picture of the $2D$ Dirac diffusing process. A Dirac bi-spinor collides with a potential step at $x=0$ with incoming momentum $\vec{p}_I$ making an angle $\theta$ with the step. Reflected and transmitted waves have momentum $\vec{p}_R$ and $\vec{p}_T$, respectively. }
\label{figd}
\end{figure}

For the potential step,
\begin{eqnarray}
V(x,y) = \begin{cases} 0 & \mbox{ for } x <0, \quad\mbox{(region $A$)}\\
V_0  & \mbox{ for } x> 0,  \quad\mbox{(region $B$)}\end{cases}
\end{eqnarray}
the solution is separated into two spatial regions as depicted in Fig.~\ref{figd}: for $x < 0$, one has the solution composed by incoming and reflected waves, and for $x >0$, one has only the transmitted wave.

Relative to the notation, the momenta are set as
\begin{equation}
{\vec{p}}_I = p \cos \theta\, \hat{x} + p \sin \theta\, \hat{y},
\end{equation}
for the incoming wave and, respectively, as
\begin{eqnarray}
\vec{p}_R &=& - p \cos \theta\, \hat{x} + p \sin \theta\, \hat{y}, \\
{\vec{p}_T} &=& q \cos \theta^\prime\, \hat{x} + q \sin \theta^\prime\, \hat{y},
\end{eqnarray}
for reflected and transmitted waves, with $p^2 = E^2 - m^2$, $q^2 = (E - V_0)^2 - m^2$, and $\theta^\prime\,$ parameterizing the $x-y$ angle of the transmitted wave.

Denoting the amplitudes by $I_\pm$, for the incoming plane waves with helicity eigenvalues corresponding to $h =\pm \, 1$, and by $R_\pm$ and $T_\pm$, respectively, for reflected and transmitted plane waves also with $h =\pm \, 1$, the bi-spinor structure of the corresponding Dirac plane wave solutions can be written in terms of
\begin{subequations}
\label{psi01}
\begin{eqnarray}
\vert \psi_I (x,y,t) \rangle &=& ( \, I_+ \, \vert \psi_+ (E, m , \theta) \rangle + I_- \, \vert \psi_-(E, m , \theta\, ) \rangle \,) \exp[- i( E t - \vec{p}_I \cdot \vec{x})], \\
\vert \psi_R (x,y,t)\rangle &=& ( \, R_+ \, \vert \psi_+ (E, m , \pi - \theta) \rangle + R_- \, \vert \psi_-(E, m , \pi - \theta) \rangle \,) \nonumber \\ &\times& \exp[- i( E t - \vec{p}_R \cdot \vec{x})], \\
\vert \psi_T (x,y,t) \rangle &=& (\, T_+ \,\vert  \psi_+ (E- V_0, m , \theta^\prime) \rangle + T_- \, \vert \psi_- (E - V_0, m , \theta^\prime) \rangle \,)  \nonumber \\ &\times& \exp[- i( (E - V_0 )t - \vec{p}_T \cdot \vec{x})],
\end{eqnarray}
\end{subequations}
with
\small
\begin{equation}
\vert \psi_{\pm} (E,m, \theta) \rangle =  \sqrt{\frac{E - m}{4 E}} \, \left[ \vert 1 \rangle \otimes( \, \pm \vert + \rangle + e^{i \theta} \vert - \rangle \,) + \sqrt{\frac{E+m}{E-m}} \, \vert 0 \rangle \otimes (\, \vert + \rangle \pm e^{i \theta} \vert - \rangle\,) \right].
\end{equation}
\normalsize

The discontinuity of the potential along the $x$-axis implies that $q \sin \theta^\prime\, = p \sin \theta\,$ such that the transmission angle can be implicitly written as the relativistic Snell's law,
\begin{equation}
\sin \theta^\prime\, = \sin \theta\, \, \sqrt{\frac{E^2 - m^2}{(E - V_0)^2 - m^2}} = \sin \theta\, \, \sqrt{\frac{1 - \mu^2}{(1 - \nu)^2 - \mu^2}},
\end{equation}
where $\mu = m/E$, $\nu = V_0/E$, and $0 < \theta < \pi/2$.
At first glance, one could separate the incident energy zones into two pieces, one for $\nu \leq 1$, which corresponds to the particle's diffusion regime, and another for $\nu > 1$, which corresponds to the Klein zone regime.
However, if one considers that the transmitted waves could exhibit oscillatory and evanescent behaviors, it would be more convenient to divide the problem into three energy zones depending on the values of the parameters $\mu$ and $\nu$. In region B above the barrier, the transmitted bi-spinor wave, $\psi_T$, is oscillatory if
\begin{eqnarray}
1 > \nu + \sqrt{(1 - \mu^2) \sin^2 \hspace{-.1cm}\theta\, + \mu^2} \hspace{2.5 cm} \mbox{(diffusion zone)}, \nonumber \\
\mu < 1 < \nu - \sqrt{(1 - \mu^2) \sin^2 \hspace{-.1cm}\theta\, + \mu^2} \hspace{2.5 cm} \mbox{(Klein zone)},
\end{eqnarray}
or evanescent if
\begin{equation}
\vert 1 - \nu \vert < \sqrt{(1 - \mu^2) \sin^2 \hspace{-.1cm}\theta\, + \mu^2} \hspace{1.5 cm} \mbox{(tunelling zone)}
\end{equation}
which can also be separated into two pieces, $\nu \leq 1$ and $\nu > 1$, as mentioned above.

The Klein zone exhibits the Klein paradox \cite{n019} in which the reflection probability is larger than the incidence probability.
Such an excess of particle number is compensated by the production of an antiparticle number by the transmitted wave at the step potential \cite{klein01,klein02}. The oscillatory behavior of the transmitted wave for both particle's diffusion and Klein zone regimes occurs for incidence angles such that
\begin{equation}
\label{007}
\sin^2 \hspace{-.1cm}\theta\, < \frac{(1 - \nu)^2 - \mu^2}{1 - \mu^2} = \sin^2 \hspace{-.1cm}\theta_{c}.
\end{equation}
For angles $\theta\, > \theta_{c}$, the transmitted wave is suppressed by an evanescent behavior as to have $\theta_{c}$ analogous to the critical angle of the Snell's law.

Once the optical parameters are established, the complete solutions given in terms of Dirac bi-spinor components, $\psi_I$, $\psi_R$ and $\psi_T$ from Eqs.~(\ref{psi01}), are resumed by
\begin{equation}
\psi_A (x,y,t) = \psi_I (x,y,t) + \psi_R (x,y,t),
\end{equation}
for region $A$, and
\begin{equation}
\psi _{B}(x,y,t) = \psi_T (x,y,t),
\end{equation}
for region $B$, satisfying the boundary (continuity) condition set as
\begin{equation}
\psi_A (0,y,t) = \psi_B (0,y,t),
\end{equation}
which implies that,
\begin{subequations}
    \begin{eqnarray}
    \label{eqscorr}
    I_+  -  I_-  + R_+  - R_- &=& \sqrt{\frac{1 - \nu - \mu}{(1 - \nu)(1 - \mu)}}\,( T_+  - T_-), \\
    e^{i (\theta\, - \theta^\prime)}(I_+ + I_-) - e^{-i(\theta\, + \theta^\prime)}(R_+ + R_-) &=& \sqrt{\frac{1 - \nu - \mu}{(1 - \nu)(1 - \mu)}}\,( T_+ + T_- ), \\
    I_+ + I_- + R_+ + R_- &=& \sqrt{\frac{1 - \nu + \mu}{(1 - \nu)(1 + \mu)}}\, (T_+ + T_-), \\
    e^{i (\theta\, - \theta^\prime)}(I_+ - I_-) -  e^{-i(\theta\, + \theta^\prime)}(R_+ - R_-) &=& \sqrt{\frac{1 - \nu + \mu}{(1 - \nu)(1 + \mu)}} \, (T_+ - T_-).
    \label{eqscorr2}
    \end{eqnarray}
\end{subequations}

The explicit expressions for $R_\pm$ and $T_\pm$ can be computed through
\begin{eqnarray}
R_\pm &=& \pm i \, \mbox{Im}[A] \,I_\pm \mp \mbox{Re}[A] \, I_\mp,  \nonumber\\
T_\pm &=& \mbox{Re} \left[ e^{i \, \frac{\theta\, - \theta^\prime\,}{2}} +  e^{i \, \frac{\theta\, + \theta^\prime\,}{2}} \, A \right] \, I_\pm + i \, \mbox{Im} \left[e^{i \, \frac{\theta\, - \theta^\prime\,}{2}} -  e^{i \, \frac{\theta\, + \theta^\prime\,}{2}} \, A \right] \, I_\mp,
\label{eqscorr3}
\end{eqnarray}
with $A$ written in terms of
\begin{equation}
\label{003}
A = \frac{(\mu \cos \theta\, + i \sin \theta)\, \cos^2 \hspace{-.1cm}\theta_{c}}{\cos^2 \hspace{-.1cm}\theta_{c} - (1 \pm \sqrt{\mu^2 \cos^2 \hspace{-.1cm}\theta_{c} + \sin^2 \hspace{-.1cm}\theta_{c}} ) \, (\, \cos^2 \hspace{-.1cm}\theta\, + \cos \theta\, \sqrt{\sin^2 \hspace{-.1cm}\theta_{c} - \sin^2 \hspace{-.1cm}\theta} )},
\end{equation}
where the $+$ sign in the denominator corresponds to the results for the diffusion zone, and the $-$ sign to the results for to the Klein zone.
The results from Eqs.~(\ref{eqscorr}-\ref{eqscorr2}) and (\ref{eqscorr3}) have been manipulated for (oscillatory and evanescent tunneling) diffusion zone(s).
By changing $1-\nu$ into $\nu-1$, one straightforwardly obtains the corresponding results for the (oscillatory and evanescent) Klein zone(s).
For a normalized incoming wave, one notices that $\vert I_+ \vert^2 + \vert I_- \vert^2 = 1 $, and that the probability conservation yields
\begin{equation}
\label{006}
\vert R_+ \vert^2 + \vert R_- \vert^2 + \frac{v_{q,x}}{v_{p,x}} \, \left[\vert T_+ \vert^2 + \vert T_- \vert^2 \right] = 1,
\end{equation}
where $v_{q,x}$ and $v_{p,x}$ are the $x$ component of the particle velocities, i. e.
\begin{eqnarray}
v_{q,x} &=& v_q \cos \theta^\prime\, = \frac{q}{E - V} \cos \theta^\prime\, = \cos \theta^\prime\, \sqrt{\frac{(1 - \nu)^2 - \mu^2}{(1 - \nu)^2}},  \nonumber \\
v_{p,x} &=& v_p \cos \theta\, = \frac{p}{E} \cos \theta\, = \cos \theta\, \sqrt{1 - \mu^2}.
\end{eqnarray}

For $\sin^2 \hspace{-.1cm}\theta_{c} = 1$ there is no potential barrier. In this case, $\vert R_+ \vert^2 = \vert R_- \vert^2 = 0$ .
Likewise, when the incidence angle reaches the value of the critical angle, i. e. $\theta\, \to \theta_{c}$, the potential step region $B$ that drives oscillating waves turns into a barrier penetration energy zone for evanescent waves.
Fig.~\ref{fig:01} depicts the curves for reflection and transmission probabilities, respectively, $\vert R_+ \vert^2 + \vert R_- \vert^2$ and $\frac{v_{q,x}}{v_{p,x}} \, \left[\vert T_+ \vert^2 + \vert T_- \vert^2 \right]$, in both diffusion and Klein zones.
The results are given in terms of the incidence angle through ($\sin \theta$) for three different values of $\sin^2 \hspace{-.1cm}\theta_{c}$, from which one can notice the transition between diffusing and tunneling behaviors for $\theta = \theta_c$.
The negative values of $\frac{v_{q,x}}{v_{p,x}} \, \left[\vert T_+ \vert^2 + \vert T_- \vert^2 \right]$ in the Klein zone corresponds to the production of an antiparticle number in order to compensate the growing behavior of $\vert R_+ \vert^2 + \vert R_- \vert^2$ so as to be consistent with continuity equation from (\ref{006}).

\section{Entanglement between helicity and intrinsic parity}

The Kronecker product structure of solution (\ref{001}) allows one to interpret Dirac quantum states as describing a composed quantum system having two internal DoF: one associated to the intrinsic parity, $P$, and another associated to the helicity quantum number, $h$.
The entanglement between these two DoF corresponds to a quantum correlation that measures the separability between spin and intrinsic parity subsystems - if such Dirac bi-spinor structure is not separable, it is due to the {\em parity-spin} ($P - h$) entanglement \cite{n020}.

The density matrix associated to each one of the states $\alpha = I$, $R$ or $T$ can be written as
\begin{equation}
\label{004}
\rho_\alpha = \vert \psi_\alpha \rangle \langle \psi_\alpha \vert,
\end{equation}
with the bi-spinor structure of $I$, $R$ and $T$ waves summarized by
\begin{equation}
\vert \psi_\alpha (x,y,t)  \rangle = \vert \psi_\alpha \rangle\,\,\exp[-i (E t - \vec{p}_\alpha \cdot \vec{x})],
\end{equation}
such that,
\begin{equation}
\vert \psi_\alpha \rangle =  \frac{1}{\sqrt{1 + \kappa_\alpha ^2}} \, [\, \vert 1 \rangle  \otimes  (\alpha_+ \vert h_{\alpha,+} \rangle \, + \, \alpha_- \vert h_{\alpha,-} \rangle)\, + \, \kappa_\alpha   \vert 0 \rangle  \otimes (\alpha_+ \vert h_{\alpha,+} \rangle \, - \, \alpha_- \vert h_{\alpha,-} \rangle) \,],
\end{equation}
where
\begin{equation}
\vert h_{\alpha,\pm} \rangle = \frac{1 \pm \vec{\sigma}\cdot \hat{p}_{\alpha}}{\sqrt{2}} \, \vert \pm \rangle,
\end{equation}
and
\begin{eqnarray}
\kappa_I = \kappa_R &=& \sqrt{\frac{1 - \mu}{1 + \mu}},\quad \mbox{and}\quad
\kappa_T = \sqrt{\frac{1 - \nu - \mu}{1 -\nu + \mu}},
\label{kappa}
\end{eqnarray}
with $\nu$ constrained to values given in terms of $\sin \theta_{c}$ and $\mu$ through Eq.~(\ref{007}).

As mentioned above, the state $\rho_\alpha$ describes a composed system of two DoF in the composite Hilbert space $\mathcal{H}=\mathcal{H}_P \otimes \mathcal{H}_h$, for which the trace of (\ref{004}) is computed through
\begin{equation}
\mbox{Tr}[\rho_\alpha] = \vert \alpha_+ \vert ^2 + \vert \alpha_- \vert^2,
\end{equation}
and from which one notices that $\rho_R$ and $\rho_T$ are not normalized and satisfy the relation
\begin{equation}
\mbox{Tr}[\rho_R] +\frac{v_{q,x}}{v_{p,x}} \, \mbox{Tr}[\rho_T] = 1.
\end{equation}
Entanglement shall be given in terms of the functionals of the normalized density operator $\rho_\alpha ^N$ as to have
\begin{equation}
\label{e002}
\rho_\alpha ^N = \frac{\rho_\alpha}{\mbox{Tr}[\rho_\alpha]} = \frac{\rho_\alpha}{\vert \alpha_+ \vert ^2 + \vert \alpha_- \vert^2}.
\end{equation}
The probability of measuring {\em odd} and {\em even} parity projections from  the state (\ref{e002}), $P_{\alpha, 1}$ and $P_{\alpha, 0}$, are calculated through
\begin{subequations}
\label{e003}
\begin{eqnarray}
P_{\alpha, 1} &=& \mbox{Tr}_{P}[\vert 1 \rangle \langle 1 \vert \, \rho_{\alpha,P} ^N \,] = \frac{1}{1 + \kappa_\alpha ^2}, \\
P_{\alpha, 0} &=& \mbox{Tr}_{P}[\vert 0 \rangle \langle 0 \vert \, \rho_{\alpha,P} ^N \,] = \frac{\kappa_\alpha ^2}{1 + \kappa_\alpha ^2},
\end{eqnarray}
\end{subequations}
where $\rho_{\alpha, P} ^N = \mbox{Tr}_h[\rho_\alpha ^N]$ is the density matrix reduced to parity space.
Probabilities (\ref{e003}) depend only on the parameters $\mu$ and $\nu$ as through Eq.~(\ref{kappa}), and are not affected by the incidence angle, $\theta$, in the barrier diffusion process. By the way, the average parity is
\begin{equation}
\langle P \rangle_\alpha = \mbox{Tr}_P [\mathcal{P} \rho_{\alpha,P} ^N] = \frac{1 - \kappa_\alpha ^2}{1 + \kappa_\alpha ^2},
\end{equation}
with $\mathcal{P} = \mbox{Diag}\{ +1 , +1 , -1, -1 \}$ the parity operator. As the previous probabilities, the mean parity does not depend on the details of the scattering process, namely on $\theta$ or $\theta^{\prime}$.

As $\rho_\alpha ^N$ is a pure state, it only admits a unique Schmidt decomposition that is given in terms of the basis vectors of each Hilbert space \cite{n021}, $\mathcal{H}_{P}$, for parity, and $\mathcal{H}_{h}$, for helicity, and of the Schmidt coefficients of the pure state decomposition.
If it exhibits some level of entanglement between $P$ and $h$ DoF, then the reduced density matrix of both of them, either $\rho_p$ and $\rho_h$ results into mixed states. One infers that a pure state is entangled if the two of its reduced systems are mixed.
The entropy of entanglement between two subsystems that compose the state $\rho_\alpha ^N$ is therefore defined by
$S_\alpha = S[\rho_{\alpha,P}] = S[\rho_{\alpha,h}] = - \sum \lambda_k \log{\lambda_k}$ \cite{n021},
where $S[\rho]$ is the vN entropy of the corresponding density matrix which is computed in terms of the $\rho$ eigenvalues, $\lambda_k$.
Pure states in the reduced representation set the entropy of entanglement equals to zero.
It implies that the state $\rho_\alpha ^N$ is separable.
Likewise, reduced mixed states lead to non-vanishing values to the vN entropy, which implies that $\rho_\alpha ^N$ is entangled.

\begin{figure}[b]
\includegraphics[width = 8 cm]{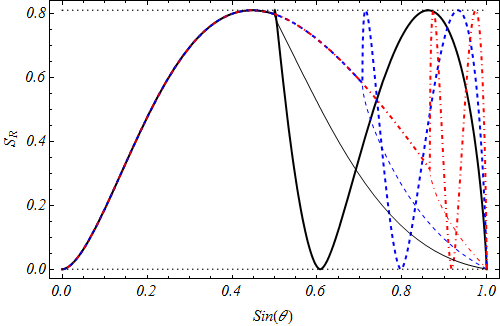}
\includegraphics[width = 8 cm]{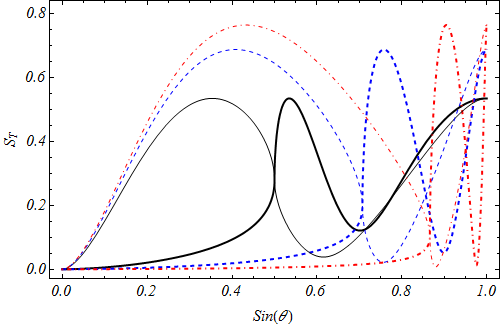}
\caption{Quantum entropies (entanglement), $S_R$ (first plot), and $S_T$ (second plot), as functions of $\sin \theta$.
Again, the plots are for $\mu = 0.5$, $I_+ = 1$ and  $\sin\theta_{c} = 1/2$ (continuous line),  $\sin\theta_{c} = 1/\sqrt{2}$ (dashed line) and  $\sin\theta_{c} = \sqrt{3}/2$ (dot-dashed line) for diffusion (thick lines) and Klein (thin lines) energy zones.
One notices the discontinuity of the entanglement quantifier derivative at $\sin{\theta} = \sin{\theta_{c}}$ indicating the transition between diffusion (c. f. oscillatory behavior) and barrier penetration (c. f. evanescent behavior) zones.
The oscillatory regime shows coincident results for reflected waves in both diffusion and Klein zones.
The two horizontal dotted lines correspond to the minimum and maximum values of the entropy given by $S_R$.}
\label{fig:02}
\end{figure}

The reduced representation of $\rho_\alpha$ to intrinsic parity subsystem $\rho_{\alpha,P} = \mbox{Tr}_h[\rho_\alpha ^N]$ has eigenvalues
\begin{equation}
\lambda_{\alpha,\pm} = \frac{1}{2} \pm \sqrt{\frac{1}{4} - 4 \frac{\kappa_{\alpha}^2}{(1 + \kappa_\alpha ^2)^2} \, \frac{\vert \alpha_+ \vert^2 \, \vert \alpha_- \vert^2}{(\vert \alpha_+ \vert^2 + \vert \alpha_- \vert^2)^2}}.
\label{00222}
\end{equation}
The entropy of entanglement is therefore given by
\begin{equation}
\label{002}
S_{\alpha} = - \lambda_{\alpha,+} \log_2 \lambda_{\alpha,+} - \lambda_{\alpha,-} \log_2 \lambda_{\alpha,-}.
\end{equation}
Fig.~\ref{fig:02} shows the plots for the vN quantum entropies for reflected and transmitted waves, $S_R$ and $S_T$, as function of the incidence angle, $\theta$, in terms of $\sin \theta\,$, for the same values of $\sin^2 \hspace{-.1cm}\theta_{c}$ as depicted in Fig.~\ref{fig:01}.
One can notice the discontinuity of the entanglement quantifier derivative at $\sin{\theta} = \sin{\theta_{c}}$, which sets the point for the transition between diffusion and barrier penetration zones.
It is important to emphasize that, for oscillatory regimes, i.e. for $\theta < \theta_c$, the $S_R$ plots return the same profiles, which does not depend on the choice of $\theta_c$, and all the corresponding curves are superposed.
Once the transition to the evanescent occurs, one has $\theta > \theta_c$.
By taking a look at the denominator of $A$ from Eq.~(\ref{003}), one notices that $\sqrt{\sin^2 \hspace{-.1 cm} \theta_c - \sin^2 \hspace{-.1 cm} \theta }$ is then automatically converted into $i \, \sqrt{\sin^2 \hspace{-.1 cm} \theta - \sin^2 \hspace{-.1 cm} \theta_c } $, i. e.  the real and imaginary parts of $A$ are changed such in a way that the $S_R$ plots shall return different profiles for the barrier penetration zones.
Given that such a transformation of real(imaginary) into imaginary(real) parts of $A$ is not analytic, it results into a discontinuity in the derivatives of the curves obtained from Eqs.~(\ref{00222}-\ref{002}) for the entanglement quantifier.

Finally, these results can also be extended to the discussion of statistical mixtures as $\rho_I = I_+ \vert \psi_+ \rangle \langle \psi_+ \vert +  I_- \vert \psi_- \rangle \langle \psi_- \vert$, for which, the entanglement could be computed either in terms of concurrence \cite{n024} or in terms of logarithmic negativity \cite{n025}. Our preliminary calculations have revealed that, for maximally mixed incident states ($I_+ = I_- = 1/2$), the entanglement for both incident and reflected states cancels off and vanishes.

\subsection{Correspondence to the averaged chirality}

\begin{figure}[b]
\includegraphics[width = 7.5 cm]{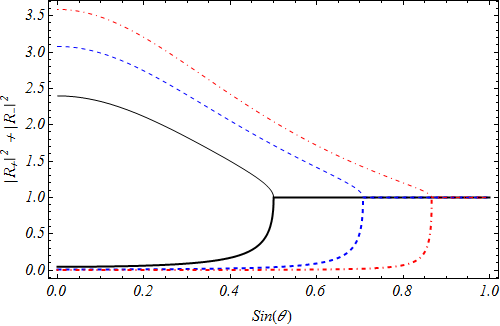}
\includegraphics[width = 8 cm]{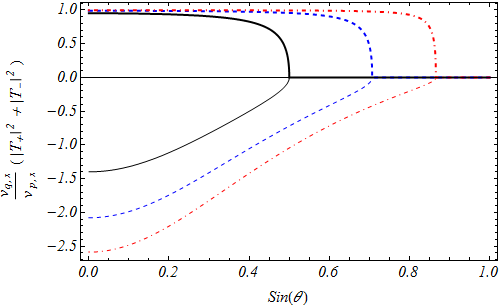}
\caption{Reflection (first plot) and transmission squared amplitudes (second plot) for diffusion (thick lines) and Klein (thin lines) energy zones, namely $\nu \leq 1$ and $\nu > 1$, as function of $\sin \theta$ for $\mu = 0.5$, $I_+ = 1$ and $\sin\theta_{c} = 1/2$ (continuous line),  $\sin\theta_{c} = 1/\sqrt{2}$ (dashed line) and  $\sin\theta_{c} = \sqrt{3}/2$ (dot-dashed line).
$\theta = \theta_c$ indicates the transition between oscillatory and evanescent behaviors and corresponds to a saturation point in the evanescent regime, for which $|T|= 0$ and $|R| = 1$.}
\label{fig:01}
\end{figure}

\begin{figure}[b]
\includegraphics[width = 8 cm]{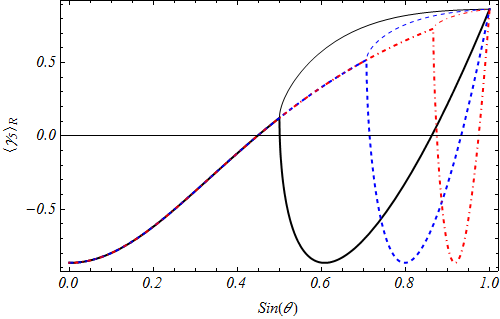}
\includegraphics[width = 8 cm]{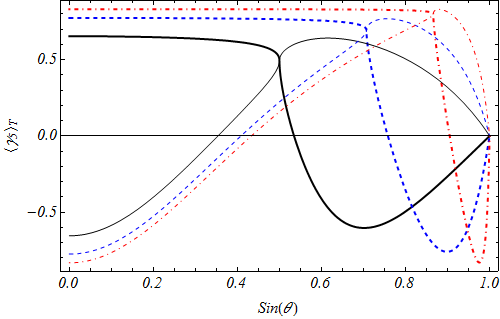}
\caption{The averaged value of the chiral operator $\langle\gamma^{\5}\rangle$ as function of $\sin \theta$, for the reflected (first plot) and transmitted (second plot) waves for diffusion (thick lines) and Klein (thin lines) energy zones, and with
$\sin\theta_{c} = 1/2$ (continuous line),  $\sin\theta_{c} = 1/\sqrt{2}$ (dashed line) and  $\sin\theta_{c} = \sqrt{3}/2$ (dot-dashed line), in correspondence to Fig.~\ref{fig:01}.
Again, $\theta = \theta_c$ indicates a (discontinuous derivative) transition between oscillatory and evanescent behaviors.}
\label{figchir}
\end{figure}

For completeness, the chiral profile of the Dirac bi-spinors can be identified through the averaged value of the chiral operator, $\langle \gamma^{5}\rangle$, with $\gamma^{5}$ written in the two qubit $SU(2) \otimes SU(2)$ representation as $\gamma^5 \equiv \sigma_x^{(1)} \otimes I^{(2)}_2$. Explicit calculations yields
\begin{equation}
\langle \gamma^5 \rangle_I = \sqrt{1 - \mu^2}\left( \vert I_+ \vert^2 - \vert I_- \vert^2  \right),
\end{equation}
for the incident wave and
\begin{eqnarray}
\langle \gamma^5 \rangle_R &=& \sqrt{1 - \mu^2}\left( \vert R_+ \vert^2 - \vert R_- \vert^2  \right), \nonumber \\
\langle \gamma^5 \rangle_T &=& \pm \sqrt{\frac{(1 - \mu^2) \sin^2  \hspace{-.1cm} \theta_c}{(1 - \mu^2) \sin^2  \hspace{-.1cm} \theta_c + \mu^2}}\left( \vert T_+ \vert^2 - \vert T_- \vert^2  \right),
\end{eqnarray}
for reflected and transmitted waves, with $+$ and $-$ signs in the expression for $\langle \gamma^5 \rangle_T$ corresponding to diffusion and Klein zones. Fig.~\ref{figchir} shows the behavior of $\langle \gamma^{5}\rangle_{R,T}$  as function of $\sin \theta$ for the same parameters of Fig.~\ref{fig:01}.
From Fig.~\ref{fig:01}, one can identify a very nice qualitative correspondence between the averaged chirality and the vN quantum entropy from Fig.~\ref{fig:02}: the results for reflected waves in the oscillatory regime do not depend on the choice of the critical angle, $\theta_c$, in the sense that the plots are all coincident; the transition to the barrier penetration (evanescent) behavior is accurately quantified by $\theta = \theta_c$; at least qualitatively, in both diffusion and Klein zone regimes, the plots indicate the possibility of there being some (quantum) correlation between $\langle \gamma^{5}\rangle_{R,T}$ and $S_{R,T}$.

\subsection{Extremal points of the entanglement}

In the diffusion and Klein zones, it is possible to analytically compute the extremal points that maximize and minimize the entanglement for the reflected states. The extremal points satisfy the condition given by
\begin{equation}
\frac{\partial S_R}{\partial \theta} \bigg{\vert}_{\theta\, = \theta_0} = 0,
\end{equation}
which implies that
\begin{equation}
\frac{\partial}{\partial \theta} \left[ \frac{\vert R_+ \vert^2 \vert R_- \vert^2}{(\vert R_+ \vert^2 + \vert R_- \vert^2)^2} \right] \bigg{\vert}_{\theta\, = \theta_0} = 0.
\end{equation}
If $I_{\pm}$ is real, the above equation has two solutions
\begin{eqnarray}
\label{005}
\sin \theta_{0} ^{(1)} &=& 0, \nonumber \\
\sin \theta_{0} ^{(2)} &=& \frac{\mu}{\sqrt{1 + \mu^2}}.
\end{eqnarray}
The corresponding eigenvalues of $\rho_{R,P}$ read
\begin{eqnarray}
\lambda_{\pm} ^{(1)} &=& \frac{1}{2} \pm \frac{1}{2} \sqrt{1 - 4 (1 - \mu^2)\vert I_+ \, I_- \vert^2}, \nonumber \\
\lambda_{\pm}^{(2)} &=& \frac{1 \pm \mu}{2},
\end{eqnarray}
from which the entropy of entanglement is calculated through ($\ref{002}$).
In spite of not obtaining analytical expressions for the extremal points of the transmitted wave entanglement, one can identify the extremal points for the quantum entropy from the plots depicted in Fig.~\ref{fig:02}.

\subsection{Relativistic and non-relativistic limits}

The ultra-relativistic limit of the above obtained results are computed through the limit of $\mu \rightarrow 0$, for which one has
\begin{equation}
A(\mu \rightarrow 0) = A_{UR} =\frac{ i \, \sin \theta\, \, \cos^2 \hspace{-.1cm}\theta_{c}}{\cos^2 \hspace{-.1cm}\theta_{c} - (1 \pm \sin^2 \hspace{-.1cm}\theta_{c}) (\, \cos^2 \hspace{-.1cm}\theta\, + \cos \theta\, \sqrt{\sin^2 \hspace{-.1cm}\theta_{c} - \sin^2 \hspace{-.1cm}\theta})}.
\end{equation}
and
\begin{equation}
\frac{\vert R_+ \vert^2 \, \vert R_- \vert^2}{(\vert R_+ \vert^2 + \vert R_- \vert^2)^2} =\frac{\vert T_+ \vert^2 \, \vert T_- \vert^2}{(\vert T_+ \vert^2 + \vert T_- \vert^2)^2} = \vert I_+ \vert^2 \, \vert I_- \vert^2.
\end{equation}
which indicates that, in the massless limit, the entanglement for reflected and transmitted states does not depend anymore on the incidence angle, $\theta$, and equals to the entanglement for the incident state.

Otherwise, the non-relativistic limit given by $\mu \rightarrow 1$ leads to
\begin{equation}
A(\mu \rightarrow 1) = A_{NR} = e^{i \theta}.
\end{equation}
as well as
\begin{equation}
\frac{\kappa_R ^2}{(1 + \kappa_R ^2)^2} = \frac{\kappa_T ^2}{(1 + \kappa_T ^2)^2} =  0,
\end{equation}
$\lambda_{+,R} = 1$, $\lambda_{-,R}= 0$ and the {\em parity-spin} entanglement is null for all incident, reflected and transmitted waves.

\subsection{Introducing phase differences}

\begin{figure}[b]
\includegraphics[width = 7.5 cm]{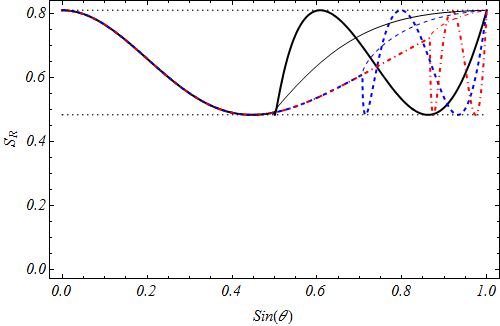}
\includegraphics[width = 7.5 cm]{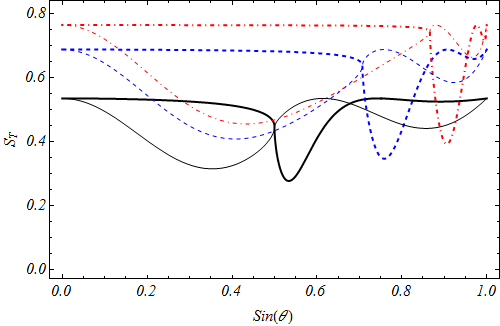}
\includegraphics[width = 7.5 cm]{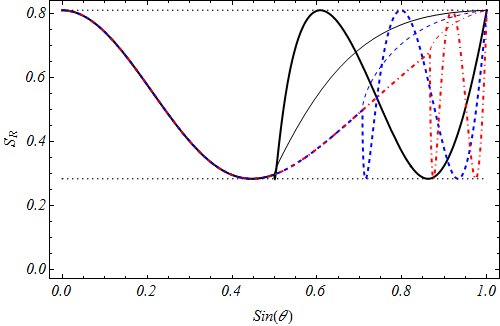}
\includegraphics[width = 7.5 cm]{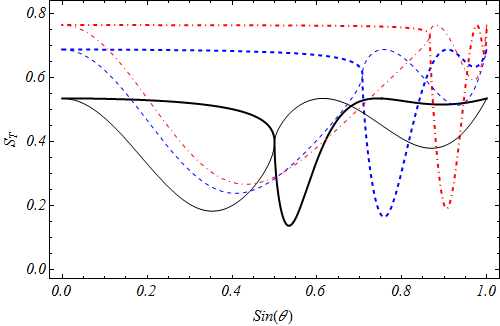}
\includegraphics[width = 7.5 cm]{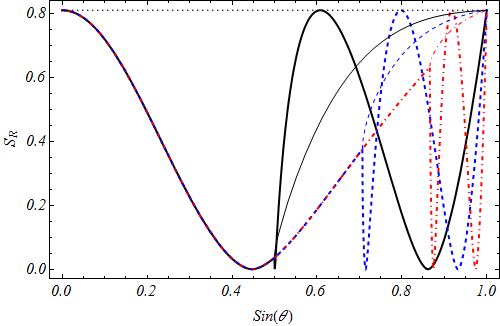}
\includegraphics[width = 7.5 cm]{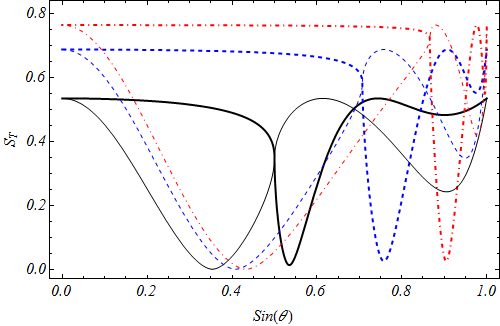}
\caption{The vN entropies, $S_R$ (first column) and $S_T$ (second column) for $\vert I_+ \vert = \vert I_- \vert = 1/\sqrt{2}$, with parameters and plot-style in correspondence with Figs.~\ref{fig:01} and \ref{fig:02}. The relative phases are given by $\Delta \omega = \pi/4$ (first row), $\Delta \omega = \pi/3$ (second row) and $\Delta \omega = \pi/2$ (third row). One notices that the local minimum for $S_T$ decreases as the phase difference decreases. For $\Delta \omega = \pi/2$, $\theta = \theta_{0}$, with $\sin \theta_0= \mu/\sqrt{1 + \mu^2}$, $S_R$ vanishes in the diffusion zone.}
\label{fig:03}
\end{figure}

Relative phases can be relevant in a large set of quantum mechanical experiments, for instance, in the context of optical interference phenomena, such as in the double slit experiments which include a transparent plate before one of the slits \cite{n022}.
For the coefficients $I_\pm$ described by complex amplitudes, one may write
\begin{equation}
I_\pm  = e^{i \, \omega_\pm} \vert I_\pm \vert,
\end{equation}
such that the relative phase of interest is written as $\Delta \omega = \omega_+ - \omega_-$. This phase has no effect over $\vert R_+ \vert^2 + \vert R_- \vert^2$ or over $\vert T_+ \vert^2 + \vert T_- \vert^2$, however, it changes the values of $\vert R_+ \vert^2 \vert R_- \vert^2$ and $\vert T_+ \vert^2 \vert T_- \vert^2$, as one can notice, for instance, from the explicit expansion of the reflection amplitudes, as given by,
\begin{eqnarray}
\vert R_+ \vert^2 \vert R_- \vert^2 &=& (\mbox{Im}[A] \mbox{Re}[A])^2 + [(\mbox{Im}[A])^2 - (\mbox{Re}[A])^2]^2 \vert I_+  I_- \vert^2 \nonumber \\&&
\qquad- 2 (\mbox{Im}[A] \mbox{Re}[A]) [(\mbox{Im}[A])^2 - (\mbox{Re}[A])^2]\, \sin \Delta \omega \vert I_+  I_- \vert(\vert I_+ \vert^2 - \vert I_- \vert^2) \nonumber \\ &&\qquad\qquad\qquad\qquad - 4  (\mbox{Im}[A] \mbox{Re}[A])^2 \sin^2 \Delta \omega \vert I_+ I_- \vert^2,\
\label{f03}
\end{eqnarray}
Once that $\vert R_+ \vert^2 \vert R_- \vert^2$ explicitly depends upon the relative phase, it obviously affects the results for the entanglement computed through the vN entropy.
Results for $S_R$ and $S_T$ for three different values of the relative phase, $\Delta \omega$, are shown in Fig.~\ref{fig:03}, for the same set of parameters used in Figs.~\ref{fig:01} and ~\ref{fig:02}.

In particular, for a maximal superposition between positive and negative helicities obtained by setting $\vert I_+ \vert  = \vert I_- \vert= 1/ \sqrt{2}$ (c. f.  Fig.~\ref{fig:03}),
the minimal point for the quantum entropy in the diffusion zone can be analytically computed for any phase difference, as it occurs at
\begin{equation}
\sin \theta_0 = \frac{\mu}{\sqrt{1 + \mu^2}},
\end{equation}
i. e. the same minimal point found in ($\ref{004}$).

Finally, all the above results for the quantum entropies computed from Dirac bi-spinor structures related to {\em particles} (positive intrinsic parity) can be converted into similar results for {\em antiparticles} (negative intrinsic parity) : one can perform the substitutions of $E \rightarrow -E$ and $V \rightarrow - V$, and equivalently, $\mu \rightarrow - \mu$. One notices that it leads to $\mbox{Re}[A] \rightarrow -\mbox{Re}[A]$ into Eq.~(\ref{f03}) (c. f the third term of Eq.~(\ref{f03})), and the entanglement for the antiparticle state is the same as that for the particle state, with opposite phase differences. For $\Delta\omega = 0$ the results are equalized.

\section{Conclusions}

In this work, the entanglement between helicity and intrinsic parity for $2D$ projections of $3D$ plane wave solutions of the Dirac equation with bi-spinor structure has been quantified in the context of the step potential barrier problem. Results for reflection and transmission coefficients in terms of incoming helicity amplitudes, incidence angles, $\theta$, and critical angles, $\theta_{c}$, for the transition between oscillatory and barrier penetrating evanescent regimes have been obtained for impinging particles in diffusion and Klein zone energy regimes.
For pure state configurations of incident, reflected and transmitted waves, the vN quantum entropy was set as the quantum correlation quantifier for the {\em parity-spin} entanglement, which was computed as a function of $\theta$.
Extremal points have been obtained in order to show that a bi-spinor quantum correlation structure can be manipulated between minimum and maximum values through the manipulation of the geometry of an optical-like apparatus that eventually controls reflected and transmitted waves.
Still in the context of the formal framework, a very suitable qualitative correspondence between the averaged chirality, $\langle \gamma^{5}\rangle_{R,T}$, and the vN quantum entropy, $S_{R,T}$, have been identified in both diffusion and Klein zone regimes.
Figuring out the quantitative origin of such an eventual correspondence between chiral conversion and entanglement, even in an enlarged context involving Dirac equation solutions, may indicate an interesting routine to the subsequent works.

Turning our conclusions to a phenomenological view, some recent issues have considered the trapped-ion physics \cite{NJP,Casa} as a flexible platform to map several suitable effects in relativistic Dirac quantum mechanics.
Therefore, the planar diffusion and the $2D$ scattering of $SU(2) \otimes SU(2)$ bi-spinor structures may also become a useful tool for quantifying elementary quantum correlations between such relative compounding spinor substructures, or even between two arbitrary Dirac particle subsystems.
In particular, it has been shown that the $2D$ relativistic scattering for $x$-dependent potentials may exhibit some kind of entanglement between transmitted and reflected wave packets and the transverse momentum \cite{NJP}, which could also be quantified through the systematic procedure described in this paper.

Our final conclusion is that the framework presented here can be straightforwardly manipulated to compute {\em spin-spin} entanglement of nonrelativistic $2D$ systems, in particular,
either for electron-electron or electron-hole pairs in the single layer graphene or, for instance, for single trapped-ions supported by a Dirac bi-spinor dynamics, which indeed deserve a deeper investigation.

\paragraph*{Acknowledgments}
 This work was supported by the Brazilian Agency CNPq (grants 440446/2014-7, 140900/2014-4 and 300809/2013-1).


\end{document}